\documentclass[aoas,preprint]{imsart}

\RequirePackage{amsthm,amsmath,amsfonts,amssymb}
\RequirePackage[authoryear]{natbib}
\RequirePackage[colorlinks,citecolor=blue,urlcolor=blue]{hyperref}
\RequirePackage{graphicx}

\startlocaldefs
\theoremstyle{plain}

\newtheorem{ass}{Assumption}[section]
\newtheorem{theorem}{Theorem}[section]
\theoremstyle{definition}

\newtheorem{rem}{Remark}[section]

\usepackage{amssymb}
\usepackage{amsthm}

\usepackage{color}
\usepackage{bm}
\usepackage{comment}
\usepackage{hyperref}

\providecommand{\skakko}[1]{\left(#1\right)}
\providecommand{\mkakko}[1]{\left\{#1\right\}}

\newcommand{\argmax}{\mathop{\rm arg~max}\limits}

\newcommand{\abs}[1]{\left\lvert#1\right\rvert}

\endlocaldefs

\begin{document}

\begin{frontmatter}
\title{ANOVATS: A subsampling-based test to detect differences among short time series in marine studies}
\runtitle{ANOVATS: 
A test to detect differences among short time series}

\begin{aug}
\author[A]{\fnms{Yuichi}~\snm{Goto} \ead[label=e1]{yuichi.goto@math.kyushu-u.ac.jp}\orcid{0000-0002-7556-2572}},
\author[B]{\fnms{Hiroko}~\snm{Kato Solvang}\ead[label=e2]{hiroko.solvang@hi.no}\orcid{0000-0002-0330-4670}},\\
\author[C]{\fnms{Masanobu }~\snm{Taniguchi}\ead[label=e3]{taniguchi@waseda.jp}\orcid{0000-0002-4783-1894}},
\and
\author[D]{\fnms{Tone}~\snm{Falkenhaug}\ead[label=e4]{Tone.Falkenhaug@hi.no}\orcid{0000-0003-3617-7724}}
\address[A]{Faculty of Mathematics, Kyushu University\textbf{}\printead[presep={ ,\ }]{e1}}
\address[B]{Marine Mammals Research Group, Institute of Marine Research\printead[presep={,\ }]{e2}}
\address[C]{Faculty of Science and Engineering, Waseda University\printead[presep={,\ }]{e3}}
\address[D]{Plankton Research Group, Institute of Marine Research\printead[presep={,\ }]{e4}}

\end{aug}

\begin{abstract}
Assessing marine ecosystems is important for understanding the impacts of climate change and human activity, as well as for maintaining healthy oceans and ecosystems.
In marine science, it is common for biologists and geologists to identify regional differences based on expert knowledge, frequently through data visualization.
However, time series data collected through surveys in marine studies typically span only a few decades, limiting the applicability of classical time series methods. Additionally, without expert knowledge, detecting significant differences becomes challenging.
To address these issues, we introduce ANOVATS (ANOVA for small-sample time series data), a subsampling-based method to detect regional differences in small-sample time series data with a fixed number of groups.
This method bypasses the need for spectral density estimation, which requires a large number of time points in the data.
Furthermore, after detecting differences in homogeneity across all areas using the ANOVATS procedure, we devised a simple ANOVATS post hoc procedure to group the areas.
Finally, we demonstrate the effectiveness of our method by analyzing zooplankton biomass data collected in different strata of the North Sea, showing its ability to quantify differences in species between geographical areas without relying on prior biological or geographical knowledge.
\end{abstract}

\begin{keyword}
\kwd{clustering}
\kwd{climate change}
\kwd{homogeneity test}
\kwd{marine ecosystem}
\kwd{time series analysis}
\end{keyword}

\end{frontmatter}

\section{Introduction}
Recent increases in the effects of climate change, together with other human factors and environmental pressures, are accelerating changes in marine ecosystems and threatening limited natural resources. 
For example, changes in commercial fish stocks in the oceans are no longer attributable solely to catch levels; it has become necessary to consider various biological interactions (e.g. food-web) within the marine ecosystem and the environmental factors that alter them.
To address these issues, the Integrated Ecosystem Assessment (IEA) (\citealt{lev09}) was proposed as a scientific basis for ecosystem-based fishery management by National Oceanic and Atmospheric Administration (NOAA), The United Nations Educational, Scientific and Cultural Organization (UNESCO), and the International Council for the Exploration of the Sea (ICES), which consider marine resources on an international scale.
The IEA approach analyzes and synthesizes information on a wide range of ecosystem components and pressures, including natural and social scientists, stakeholders, and resource managers (\citealt{cla23}).
Based on the results analyzed, the IEA identifies the status, changes, relationships, and processes at the ecosystem level (\citealt{wki18}), assesses risks, and ultimately evaluates strategies to implement ecosystem-based management measures (\citealt{cla23}).
In particular, a working group established by scientific experts from the International Council for the Exploration of the Sea (ICES) is conducting IEAs in European waters (ICES, \url{https://www.ices.dk}).
The IEA regional group within ICES focuses on collecting time series data for key physical and biological variables and conducts integrated trend analyses to understand dynamics and ecosystem changes (\citealt{wki18}, \citealt{wki22}). 
The time series datasets considered by the IEA are collected through various scientific surveys and have been selected as indicators that form the basis for assessing the status and trends of ecosystems (ICES, \url{https://www.ices.dk}, OSPAR, \url{https://odims.ospar.org}). 

\begin{figure}[tbp]
\centering
\includegraphics[width = 12cm]{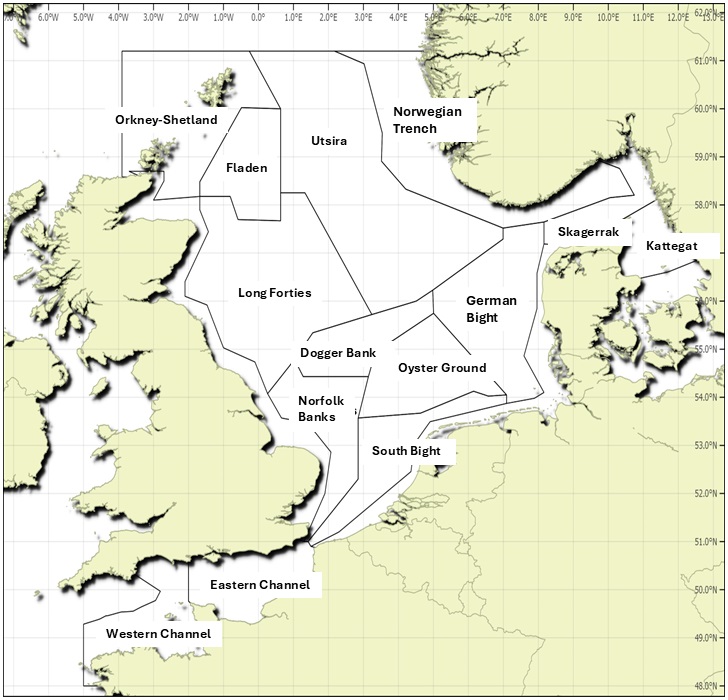}
	\vspace{-0.4cm}
		\caption{\small Fourteen strata in the North Sea considered in \citealt{ic23}. 
        This classification represents a typical approach to dividing the sea into several strata in marine studies.
        In the real data analysis in Section 4, we analyze data obtained from these areas.}
	\label{fig:strata_NS}
\end{figure}

However, there are potential challenges when analyzing changes over time.
Data have been collected for only a few decades since awareness of climate change emerged, resulting in insufficient information to apply conventional time series models that account for temporal correlations and relationships between multivariate time series (\citealt{sh19}, \citealt{wki18}, \citealt{wki22}, \citealt{ss19}). 
Additionally, the IEA is investigating differences in environmental factors across various subregions (referred to as strata (\citealt{ic23}) or polygons (\citealt{oe20})), considering the distribution of biological communities.
This is important because the assessment areas are adjacent to different oceans and possess ecologically distinct marine geological environments, which are likely to result in changes in the ecosystems.
For example, there are fourteen strata containing diverse ecosystems, ranging from the shallow waters south of the North Sea to the deep waters of the Norwegian Trench, as illustrated in Figure \ref{fig:strata_NS} (\citealt{ic23}, numerical examples in this region are presented in Section 4).
The IEA regional groups working in Norwegian waters (including Iceland, Denmark, Svalbard, the Faroe Island, and the UK, in addition to Norway) have provided trend analyses for short time series data.
This includes a method for classifying biotic (such as zooplankton or biological community) and abiotic (environmental factors) data into common trend patterns (\citealt{so23}, \citealt{ic23}).
Additionally, they have developed a method for identifying observations marked with a flag, indicating the latest observation value that is deemed to deviate from recent trends (\citealt{ic23}, \citealt{sp24}). 
Furthermore, the regional groups need consider changes in the state of the ecosystem across multiple strata (or polygons).
For this reason, analysis of variance (ANOVA) is a possible statistical method for examining the homogeneity within the strata based on the time series data.
Within ICES, as all variations and tendencies derived from time series are analysed as part of trend analysis, examining differences between strata constitutes one important task within this trend analyses.

As mentioned, the time series datasets handled by IEA are generally short, but time dependence is unavoidable. 
ANOVA was originally conceived as a method of multivariate analysis, which basically assumed identically independent distribution for the data (\citealt{searle92}, \citealt{so04}, \citealt{clarke2008}), and does not take time dependence into account. 
If a conventional analysis method that ignores time dependence is used despite the presence of actual time dependence, serious problems may arise (\citealt{br80}).
For time series data, smoothing spline ANOVA for time-dependent spectral analysis (\citealt{gu03}),
tests for the equality of several spectra (\citealt{dp09}, \citealt{jp15}),
ANOVA for time series with independent and correlated groups, studied by \cite{nt2018} and \cite{galt2022}, involved estimation of spectral density.
These approaches require time series datasets containing more than 100 time points, as seen in several textbooks on time series (e.g., \citealt{brillinger1981}, \citealt{hj19}, \citealt{shumway2006}).

To face this challenge, we propose an ANOVA test for short Time Series (ANOVATS) using a subsampling method, which makes it possible to circumvent the need for estimating the spectral density.
Estimation of the long-run variance typically requires a large sample size,
and when the sample size is insufficient, tests based on its estimation often suffer
from severe size distortion. 
To address this problem, one may consider applying resampling methods such as the bootstrap to approximate critical values.
In the frequency domain, subsampling methods are known to perform effectively
in several contexts (\citealt{mpk20, ykn23, ykn24}).
For lattice data, \citet{hl00} applied the subsampling technique to estimate
the variance of parameters of interest, thereby overcoming the difficulties
arising from replacing the unknown mean with the sample mean.
Building on these insights, we apply a subsampling method to time series data.
The idea is to divide the observed time series into shorter time series with overlap and then construct the empirical distribution of the test statistic based on the shorter time series (see \citealt{pr99}).
Subsampling-based methods have been investigated in several studies
(e.g., \citealt{c86, pr94, f99, bmp10, mpr12, dl23}),
but not in the context of ANOVA for dependent or short time series.

After rejecting the homogeneity hypothesis, Fisher's least significant difference procedure has been widely used to identify which groups exhibit significant differences. This method applies the t-test to each pair of groups, leading to a severe multiple comparisons problem \citep{h86}.
We propose a method that efficiently detects significant differences among groups while reducing the number of tests.

This paper is organized as follows: In Section 2, we introduce the one-way model with {regional differences}, time-dependent errors, and correlated groups, and we define a test statistic and a subsample-based $p$-value. The asymptotic behavior of the proposed test is investigated.
The analysis method is introduced after rejecting the null hypothesis, which helps in dividing or grouping the data into clusters.
In Section 3, we present a numerical example to demonstrate our proposed method.
In Section 4, we demonstrate our approach applied to two real data sets from the North Sea. 
In Appendix, we provide proofs of our theoretical and numerical results from a simulation study to validate the proposed approach.

\section{Statistical methodology}

\subsection{Proposed ANOVATS method}\label{sec2.1}
We consider the one-way model with {regional differences}, time-dependent errors, and correlated groups defined as
\begin{align}\label{model}
{\boldsymbol z}_{it}={\boldsymbol \mu}+{{\boldsymbol \psi}_{i}}+{\boldsymbol e}_{it}, \qquad
i=1,\ldots,a;\ t=1, \ldots,n,
\end{align}
where ${\boldsymbol z}_{it}=(z_{it1},\ldots,z_{itp})^\top$ is a $p$-dimensional observation from the $i$-th group at time $t$, ${\boldsymbol \mu}=(\mu_1,\ldots,\mu_p)^\top$ is a general mean, {${\boldsymbol \psi}_{i}=(\psi_{i1},\ldots,\psi_{ip})^\top$ is a nonrandom regional effect} such that $\sum_{i=1}^{a}{\boldsymbol \psi}_{i}={\boldsymbol 0}$, and ${\boldsymbol e}_{it}=(e_{it1},\ldots,e_{itp})^\top$ is a disturbance process such that ${\boldsymbol e}_{t}=({\bm e}_{1t}^\top,\ldots,{\bm e}_{at}^\top)^\top$ is a centered strictly stationary sequence with an $ap$-by-$ap$ spectral density matrix ${\boldsymbol f}(\lambda)=({\boldsymbol f}_{ij}(\lambda))_{i,j=1,\ldots,a}$ for $\lambda\in[-\pi,\pi]$, where ${\boldsymbol f}_{ij}(\lambda)$ is a spectral density matrix of the processes $\{{\boldsymbol e}_{it}\}$ and $\{{\boldsymbol e}_{jt}\}$. 

We are interested in the existence of {regional differences}, that is, the following hypothesis testing problem:
\begin{align}
\label{hyp}
H_0:{{{\bm\psi}_{1}}=\cdots={{\bm \psi}_{a}}}\quad\text{v.s.}
\quad H_1:\text{ $H_0$ does not hold}.
\end{align}
 For this hypothesis, we propose the test statistic defined as
\begin{align*}
T_n
&=n\sum_{i=1}^a (\overline{{\boldsymbol z}_{i.}}-\overline{{\boldsymbol z}_{..}})^\top
(\overline{{\boldsymbol z}_{i.}}-\overline{{\boldsymbol z}_{..}}).
\end{align*}
The estimation of spectral density from short time series poses a significant obstacle to achieving proper size control. To address this issue, the proposed test statistic is designed to exclude spectral estimators. As a consequence, the asymptotic variance includes information from the underlying process, making the statistic not asymptotically distribution-free. Therefore, tailored approaches are required to compute the critical values. In this paper, we overcome this limitation by employing a subsampling method. Here, we note that in classification problems under high-dimensional and low sample size settings, the inverse of the sample variance matrix in the classifier is replaced by an identity matrix or other invertible matrices, as the sample variance matrix is always singular (see, e.g., \citealt{ay14}). From this perspective, our statistic is comprehensible and natural.

We make the following assumption to establish the theoretical results:
\begin{ass}\label{as}
\begin{enumerate}
\item[(i)] The disturbance process $\{{\boldsymbol e}_{t}\}$ is geometrically $\alpha$-mixing, that is, the $\alpha$-mixing coefficient $\alpha(\cdot)$ defined as 
\begin{align*}
\alpha(n):=\sup_{k\in\mathbb Z, A\in \mathcal F_{-\infty}^k,\ B \in \mathcal F_{k+n}^{\infty}}|{\rm P}(AB)-{\rm P}(A){\rm P}(B)|,
\end{align*}
where, for $a\leq b$, $\mathcal F_{a}^b$ is the $\sigma$-field generated by $\{\bm e_t: a\leq t\leq b\}$, satisfies $\alpha(n)\leq C_\alpha\rho^n$ for some constants $C_\alpha\in(1,\infty)$ and $\rho\in(0,1)$.

\item[(ii)]
The moments of all orders exist in the sense that 
$$\sup_{s_2,s_3,\ldots,s_\ell\in\mathbb Z}\abs{{\rm E}e_{i_10d_1}e_{i_2s_2d_2}\cdots e_{i_\ell s_\ell d_\ell}}<\infty$$ for any $\ell\in\mathbb N$, any $(i_1,\ldots,i_\ell)\in\{1,\ldots,K\}^\ell$, and any $(d_1,\ldots,d_\ell)\in\{1,\ldots,p\}^\ell$.
\end{enumerate}
\end{ass}

\begin{rem}
Under Assumption  \ref{as}, it holds, for any $d\in\mathbb N$, any $\ell\in\mathbb N$, any $(i_1,\ldots,i_\ell)\in\{1,\ldots,K\}^\ell$, and any $(d_1,\ldots,d_\ell)\in\{1,\ldots,p\}^\ell$, that
\begin{align}\label{lem_summability_mixing_eq}
\sum_{s_2,\ldots,s_\ell=-\infty}^\infty\skakko{1+\sum_{j=2}^\ell\abs{s_j}^d}\abs{{\rm cum}\{e_{i_10d_1},e_{i_2s_2d_2},\ldots, e_{i_\ell s_\ell d_\ell\}}}<\infty.
\end{align}
See Lemma 2.1. of \cite{GZKC23} for this result.
\end{rem}

\begin{rem}
Note that  $\sqrt n (\overline{{\boldsymbol z}_{1.}}^\top-\overline{{\boldsymbol z}_{..}}^\top,\dots,
\overline{{\boldsymbol z}_{a.}}^\top-\overline{{\boldsymbol z}_{..}}^\top)^\top$ converges in distribution to the centered normal distribution with variance ${\boldsymbol H}$, where
${\boldsymbol H}=({\boldsymbol H}_{ij})_{i,j=1\ldots,a}
$ with 
\begin{align*}
{\boldsymbol H}_{ij}=&
2\pi{\boldsymbol f}_{ij}(0)
-\frac{2\pi}{a}\sum_{s=1}^a\mkakko{{\boldsymbol f}_{sj}(0)+{\boldsymbol f}_{is}(0)}+
\frac{2\pi}{a^2}\sum_{s,k=1}^a{\boldsymbol f}_{sk}(0),
\end{align*}
under the cumulant summability condition \eqref{lem_summability_mixing_eq}.
under the cumulant summability condition \eqref{lem_summability_mixing_eq}.
Then, under $H_0$, $T_n$ converges in distribution to $\sum_{j=1}^r \lambda_j \chi^2_{1,j}$,
where $\lambda_1 \ge \cdots \ge \lambda_r > 0$ are the nonzero eigenvalues of $\boldsymbol H$, and $\chi^2_{1,j}$ are independent $\chi^2_1$ random variables.
\end{rem}

The subsampling test statistic and the subsample-based $p$-value with block length $b$ are defined, for $t=1,\ldots,n-b+1$, as
\begin{align}\nonumber
T_{n,b,t}
&=\frac{b}{1-\frac{b}{n}}\sum_{i=1}^a (\overline{{\boldsymbol z}_{i.,b,t}}-\overline{{\boldsymbol z}_{..,b,t}})^\top
(\overline{{\boldsymbol z}_{i.,b,t}}-\overline{{\boldsymbol z}_{..,b,t}})\\\label{pvalue}
\text{ and }
p_n&=\frac{1}{n-b+1}\sum_{t=1}^{n-b+1}\mathbb I\{T_{n,b,t}> T_{n}\},
\end{align}
where 
$\overline{{\boldsymbol z}_{i.,b,t}}=\sum_{j=t}^{t+b-1}{\boldsymbol z}_{ij}/{{b}}$, 
$\overline{{\boldsymbol z}_{..,b,t}}=\sum_{i=1}^{a}\sum_{j=t}^{t+b-1}{\boldsymbol z}_{ij}/{(a{b})}$, and $\mathbb I\{\cdot\}$ is an indicator function.
The coefficient $(1-{b}/{n})^{-1}$, which is asymptotically negligible, corresponds to the finite population correction and provides improvement of the empirical size of the test. 
We propose a subsampling base test {(ANOVATS)} that rejects $H_0$ whenever $p_n<\varphi$.

This test can equivalently be expressed in the quantile-based form as
\begin{align*}
p_n = 1 - \hat F_{n,b}(T_n) < \phi
\quad &\Leftrightarrow\quad
T_n \ge \inf\{x \in \mathbb{R} : \hat F_{n,b}(x) > 1 - \phi\},
\end{align*}
where $\hat F_{n,b}(x) = (n-b+1)^{-1}\sum_{t=1}^{n-b+1}\mathbb{I}\{T_{n,b,t} \le x\}$ denotes
the empirical distribution function of the subsampling statistics.
Hence, rejecting $H_0$ when $p_n < \phi$ is equivalent to rejecting it when 
$T_n$ exceeds the empirical $(1-\phi)$-quantile of $\hat F_{n,b}$.

Note that in this paper,  we set the block length as $b=\lfloor 2.5 n^{1/3} \rfloor$ because the order $n^{1/3}$ corresponds to the optimal rate in the sense of bias-variance tradeoff for subsampling-based variance estimation of the sample mean in mixing stationary time series \cite[Section 9.2.1]{pr99}. The constant factor $2.5$ is chosen empirically as it provides good finite-sample performance (see Section A.2 in Appendix). 
For the block length $b=\lfloor 2.5 n^{1/3} \rfloor$ and  significance level $\varphi=0.05$, if there is at least one (or two) $t$ such that $T_{n,b,t}> T_{n}$, then $p_n\geq\varphi$ for $n=3,\ldots,26$ (or $48$). Thus, for small-sample time series (those with $n\leq26$), if there is at least one $t$ such that $T_{n,b,t}> T_{n}$, this immediately implies $H_0$ is not rejected.
The following theorem shows that our test has fundamental properties:

\begin{theorem}\label{thm}
Suppose Assumption \ref{as} and that the subsampling block of length $b$ satisfies $b\to\infty$ and $b/n\to0$ as $n\to\infty$. For significance level $\varphi$,
the test that rejects $H_0$  whenever $p_n<\varphi$ has  asymptotic size $\varphi$ and is consistent, that is,
\[
{\rm P}(p_n < \varphi \mid H_0)\to \varphi
\quad
\text{and}
\quad
{\rm P}(p_n < \varphi \mid H_1)\to 1
\quad
\text{as $n\to\infty$.}
\]
\end{theorem}


\subsection{Post-ANOVATS procedures}\label{sec:2.2}
Let $p=1$ and let ${\rm Area}_{i}$ represent the area we observed $\{z_{it}\}_{t=1,\ldots,n}$. In Section \ref{sec2.1}, we considered the testing method based on subsampling for  hypothesis \eqref{hyp}.
If $H_0$ is not rejected, there is no evidence for the mean-difference among areas. If $H_0$ is rejected, we conclude the area-means are not homogeneous. Then when the means are arranged in ascending order (say $\mu+{{\psi}_{[1]}},\ldots,\mu+{{\psi}_{[a]}}$), there exists at least one index $i\in\{1,\ldots,a-1\}$ such that the difference $ {{\psi}_{[i+1]}}-{{\psi}_{[i]}}$ is strictly positive. It is reasonable to divide the areas into two groups $({\rm Area}_{{[1]}},\ldots,{\rm Area}_{{[i^\prime]}})$ and $({\rm Area}_{{[i^\prime+1]}},\ldots,{\rm Area}_{{[a]}})$ for the index $i^\prime$ corresponding to one of the largest differences, that is,
$$i^\prime:=\argmax_{i=1,\ldots,a-1}({{\psi}_{[i+1]}}-{{\psi}_{[i]}}).$$
 In practice, $\mu+{\psi}_{i}$ is replaced with the sample mean $\overline{z_{i.}}$ for ${\rm Area}_i$.
Note that hierarchical splitting methods based on the sum of squares were considered by, e.g., \cite{ec65}, \cite{sk74}, \cite{cc85}, and \cite{wx14}. 

\begin{rem}
\cite{wx14} proposed a procedure for clustering data by repeatedly applying a test procedure and verified its consistency. Their idea is to take the significance level $\varphi_n$ converging to zero as $n\to\infty$ to eliminate type 1 errors. However, $\varphi_n$ must converge slow enough to ensure the power of the test tends to one. 

In our setting, for $\varphi_n$ such that $\varphi_n\to0$ and $\varphi_n/\skakko{n^{\frac{a}{2}-1}\exp\skakko{-\frac{n}{2}}} \to\infty$ as $n\to\infty$, the procedure for rejecting $H_0$ in favor of $H_1$ whenever $p_n<\varphi_n$ satisfies
\begin{align}\label{eq:phi}
{\rm P}\skakko{p_n<\varphi_n | H_0}\to0\quad
\text{and}\quad
{\rm P}\skakko{p_n<\varphi_n | H_1}\to1 \quad \text{as $n\to\infty$}.
\end{align}
The proof of the convergences in \eqref{eq:phi}  is deferred to Section B.2 in Appendix.
The derivation of the order of $\varphi_n$ is based on the upper bound of the tail of a chi-squared distribution \cite[Lemma 1]{il06}.  
\end{rem}

\subsection{Analyzing procedure}\label{sec:2.3}
Based on Sections \ref{sec:2.2} and \ref{sec:2.3}, the analyzing procedure based on ANOVATS is  summarized as follows:
\begin{enumerate}
\item[Step 1] \textbf{Hypothesis testing for all areas:}
Let $p=1$ and let ${\rm Area}_{i}$ represent the areas $\{z_{it}\}_{t=1,\ldots,n}$ that were observed. Consider the hypothesis test for the absence of a regional difference against the presence of a regional difference, defined in \eqref{hyp}.
Applying our test to the data provides the $p$-value $p_n$.
\begin{enumerate}
\item[]
If $H_0$ is not rejected ($p_n\geq0.05$), terminate the procedure.
\item[]
If $H_0$ is rejected ($p_n<0.05$), proceed to the next step.
\end{enumerate}

\item[Step 2]  \textbf{Division of areas into two groups}
After rejecting $H_0$, sort the sample means for all areas in ascending order, denoted as $\overline{z_{[1].}},\ldots,\overline{z_{[a].}}$, corresponding to the sorted areas ${\rm Area}_{[1]},\ldots,{\rm Area}_{[a]}$.
Next, compute  the differences between the sample means of adjacent areas: $\overline{z_{[i+1].}}-\overline{z_{[i].}}$ for all $i=1,\ldots,a-1$, and find the index $i^\prime$ that maximizes these differences: 
$$i^\prime:=\argmax_{i=1,\ldots,a-1}(\overline{z_{[i+1].}}-\overline{z_{[i].}}).$$

The areas are then divided into two groups:
$$\text{Group 1: ${\rm Area}_{{[1]}},\ldots,{\rm Area}_{{[i^\prime]}}$, \quad
Group 2: ${\rm Area}_{{[i^\prime+1]}},\ldots,{\rm Area}_{{[a]}}$}$$

\item[Step 3]  \textbf{Further division}  
For each of the two groups, hypothesis tests are conducted to determine if further division is necessary. The following hypotheses are considered for each group:
\begin{enumerate}
\item[] For Group 1,
$
H_0:{{{\psi}_{[1]}}=\cdots={{\psi}_{[i^\prime]}}}\quad\text{v.s.}
\quad H_1:\text{ $H_0$ does not hold.}
$

\item[] For Group 2,
$
H_0:{{{\psi}_{[i^\prime+1]}}=\cdots={{\psi}_{[a]}}}\quad\text{v.s.}
\quad H_1:\text{ $H_0$ does not hold}.
$
\end{enumerate}
The above steps are repeated for each group, testing and subdividing until the hypothesis is not rejected or the number of areas in the group is one.

\end{enumerate}

By following this procedure, the areas can be grouped into statistically significant clusters.

\begin{figure}[t]
	\begin{center}
\includegraphics[width = 12cm]{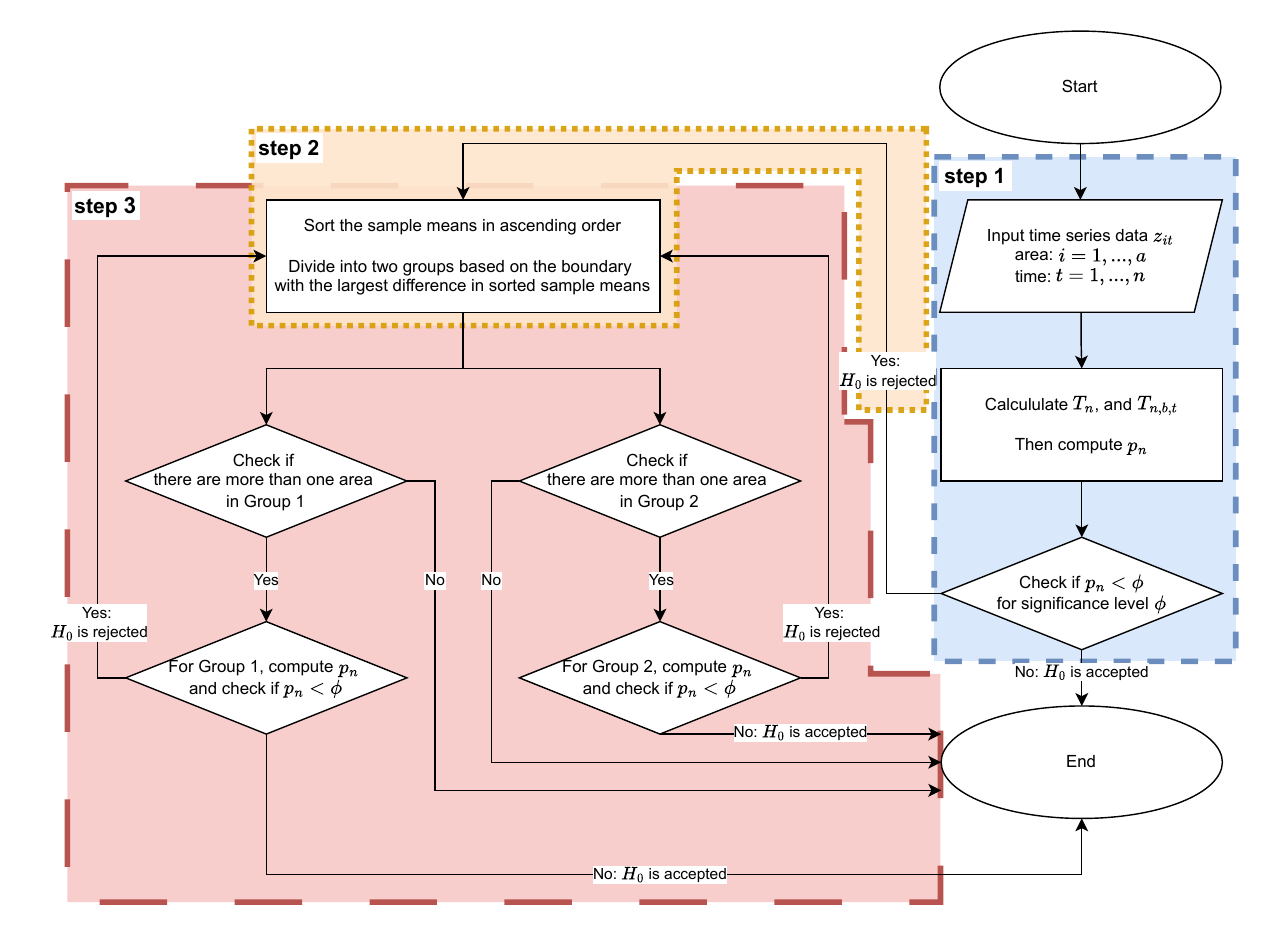}
	\end{center}
	\vspace{-0.4cm}
		\caption{\small The flowchart of the ANOVATS-based proposed procedure in Section \ref{sec:2.3}.
        {Step 1: Apply ANOVATS to the full dataset.
        Step 2: If the homogeneity hypothesis is rejected in Step 1, divide the dataset into two subgroups.
        Step 3: Repeat Steps 1 and 2 for each subgroup until the hypothesis is not rejected or each subgroup contains only one area.}}
	\label{fig:flowchart}
\end{figure}

A flowchart of this procedure is shown in Figure \ref{fig:flowchart}.
The numerical procedure is implemented using R code (\citealt{cran23}) given in Appendix.

\section{Demonstration example}

\begin{figure}[tbp]
	\begin{center}
\includegraphics[width = 12cm]{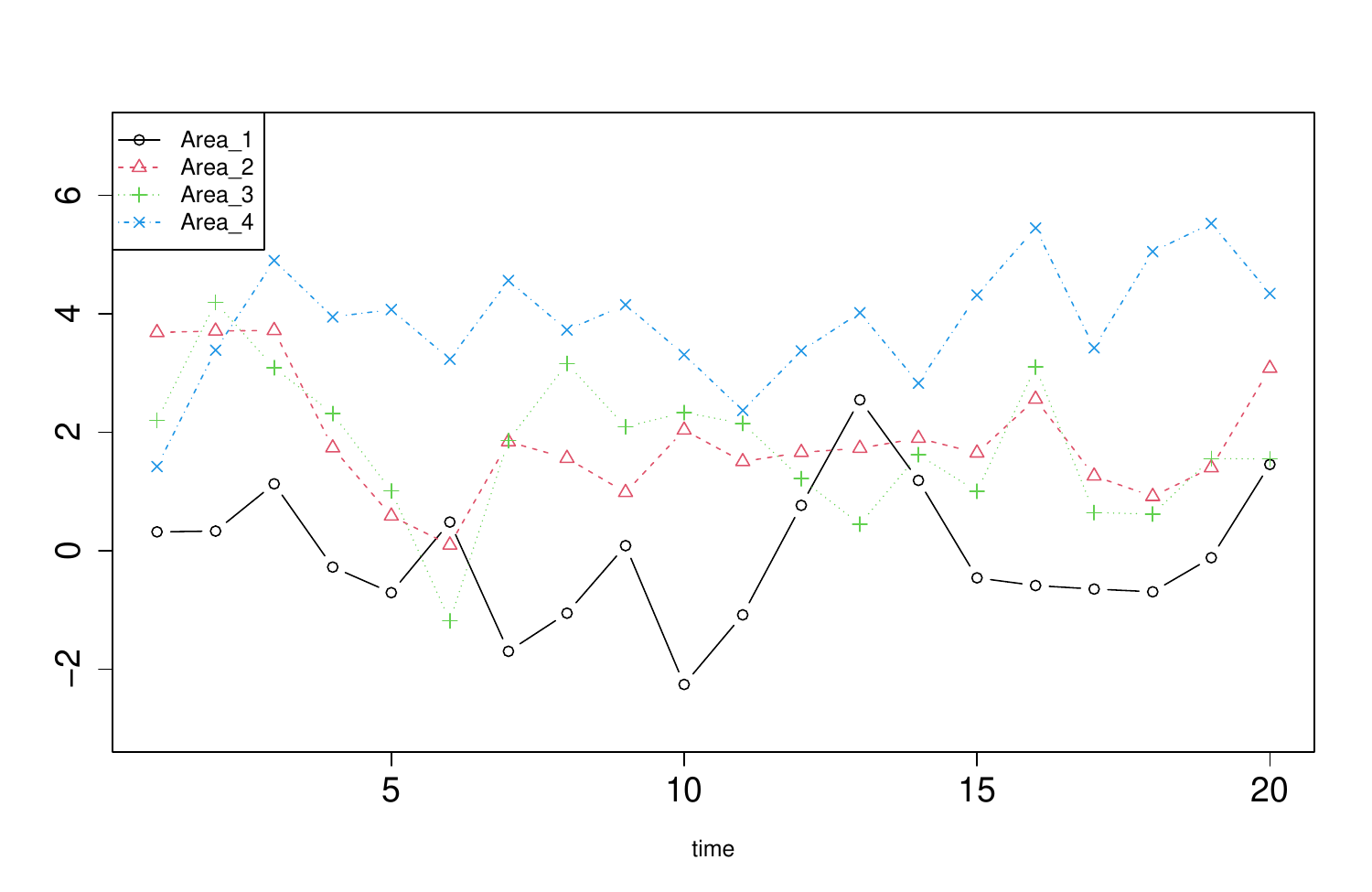}
	\end{center}
	\vspace{-0.4cm}
		\caption{\small {Plots of realizations of moving average models of order one, with means 0, 2, 2, and 4, corresponding to ${\rm Area}_1$–${\rm Area}_4$, respectively, to demonstrate our analysis methods proposed in Section \ref{sec:2.3}.}}
	\label{fig:ex1}
\end{figure}

\begin{figure}[tbp]
	\begin{center}
\includegraphics[width = 12cm]{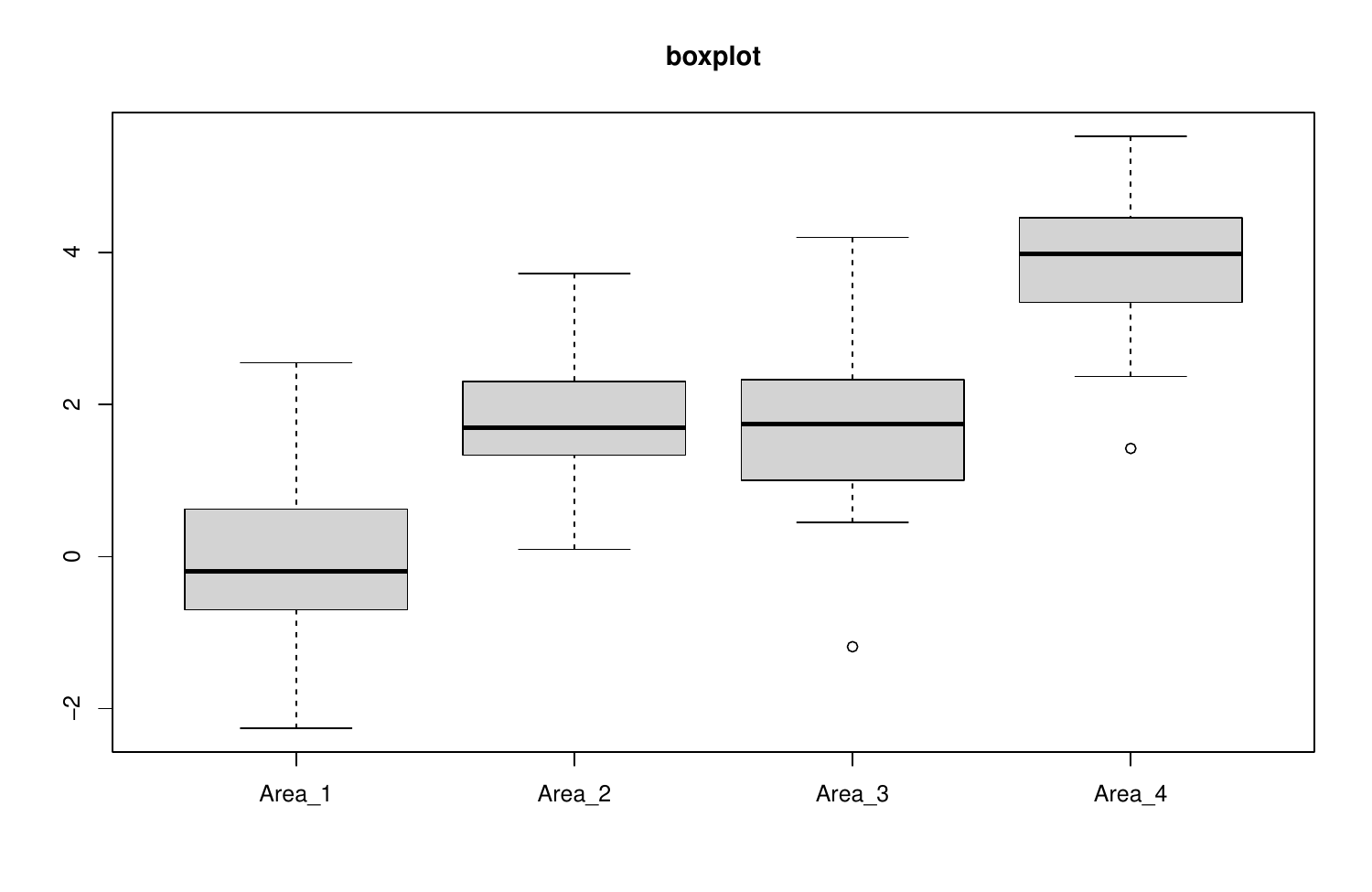}
	\end{center}
	\vspace{-0.4cm}
		\caption{\small {Boxplots of realizations of moving average models of order one, with means 0, 2, 2, and 4, corresponding to ${\rm Area}_1$–${\rm Area}_4$, respectively, to illustrate the distribution and variability of the data.}}
	\label{fig:ex2}
\end{figure}

\begin{enumerate}
\item \textbf{Model setup:}
Consider the moving average model of order 1 $y_{it}=\psi_i + e_{it}$, 
where $(\psi_1,\ldots,\psi_4)=(0,2,2,4)$, 
$e_{it}=\nu_{it} + 0.5\nu_{i(t-1)}$, and 
$\{\nu_{it}\}$ follows i.i.d.\ standard normal distribution
with $n=20$ and $a=4$. Let ${\rm Area}_i$ denote the area name where $\{z_{it}\}$ was observed.
Clearly,  we know $\psi_1<\psi_2=\psi_3<\psi_4$.
Figures \ref{fig:ex1} and \ref{fig:ex2} show the plots and boxplots of these realizations. 

\item \textbf{Initial hypothesis test:}
We performed our proposed test for the hypothesis $H_0: \psi_1=\cdots=\psi_4$ vs. $H_1: H_0$ does not hold. The test provided a $p$-value of $0$, leading to the rejection of $H_0$.

\item \textbf{Sorting areas by sample mean:} 
The sample means of $y_{1t},\ldots,y_{4t}$ are $-0.064$,  $1.88$,  $1.75$, and $3.87$, respectively. We sorted the areas in ascending order according to their sample means, resulting in the order: ${\rm Area}_1$, ${\rm Area}_3$, ${\rm Area}_2$, ${\rm Area}_4$.

\item \textbf{Calculating differences between adjacent areas:} 
We computed the differences between the sample means for each pair of adjacent rearranged areas: $$1.75-(-0.064)=1.81,\quad 1.88-1.75=0.13,\quad 3.87-1.88=1.99.$$

\item \textbf{First division of areas:}
Based on the largest difference in sample means (1.99) between ${\rm Area}_2$ and ${\rm Area}_4$, we divided the areas into two groups: the first group is (${\rm Area}_1$, ${\rm Area}_3$, ${\rm Area}_2$), and the second groups is (${\rm Area}_4$).
We stopped the procedure for the second group because it contains only one area.

\item \textbf{Hypothesis testing for the first group:} 
For the first group (${\rm Area}_1$, ${\rm Area}_3$, ${\rm Area}_2$), we consider the hypothesis $H_0: \psi_1=\psi_3=\psi_2$ vs. $H_1: H_0$ does not hold. Our test provided a $p$-value of $0$, leading to the rejection of $H_0$.

\item \textbf{Further division of the first group:}  
We then divide the first group into two subgroups based on the largest difference in sample means: the first subgroup is (${\rm Area}_1$) and the second subgroup is $({\rm Area}_3$, ${\rm Area}_2$).
We stopped the procedure for the first subgroup because it contains only one area.

\item \textbf{Final hypothesis test for the remaining areas:}  
For the second subgroup (${\rm Area}_3$, ${\rm Area}_2$), we consider the hypothesis $H_0: \psi_3=\psi_2$ vs. $H_1: H_0$ does not hold. Applying our test, we obtained a $p$-value of $0.73$, indicating that  $H_0: \psi_3=\psi_2$ cannot be rejected. 

\item \textbf{Final grouping:}  
In the end, we obtained three statistically significantly different groups (${\rm Area}_1$), (${\rm Area}_3$, ${\rm Area}_2$), and (${\rm Area}_4$).

\end{enumerate}


\section{Empirical studies}
Two numerical examples, {related to zooplankton biomass data in the North Sea}, applying our proposed ANOVA for time series data are summarized in this section. 
{As mentioned in Introduction, the North Sea includes different subregions (strata) of oceanography, as shown in Figure \ref{fig:strata_NS}. 
Zooplankton are at the bottom of the food chain and have an impact on the marine ecosystem of fish. However, owing to recent climate change, there is debate about changes in the currents and biological factors of the North Sea (\citealt{haj02}), and existing knowledge alone is insufficient to interpret changes in biomass. 
We demonstrates the following examples using ANOVATS, which conduct systematically and statistically evaluateing differences in plankton biomass in strata of the North Sea without biological or geological knowledge as prior information.
}

\subsection{Zooplankton data in the North Sea}\label{Zplk_NS}
\begin{figure}[htb]
	\begin{center}
\includegraphics[width = 12cm]{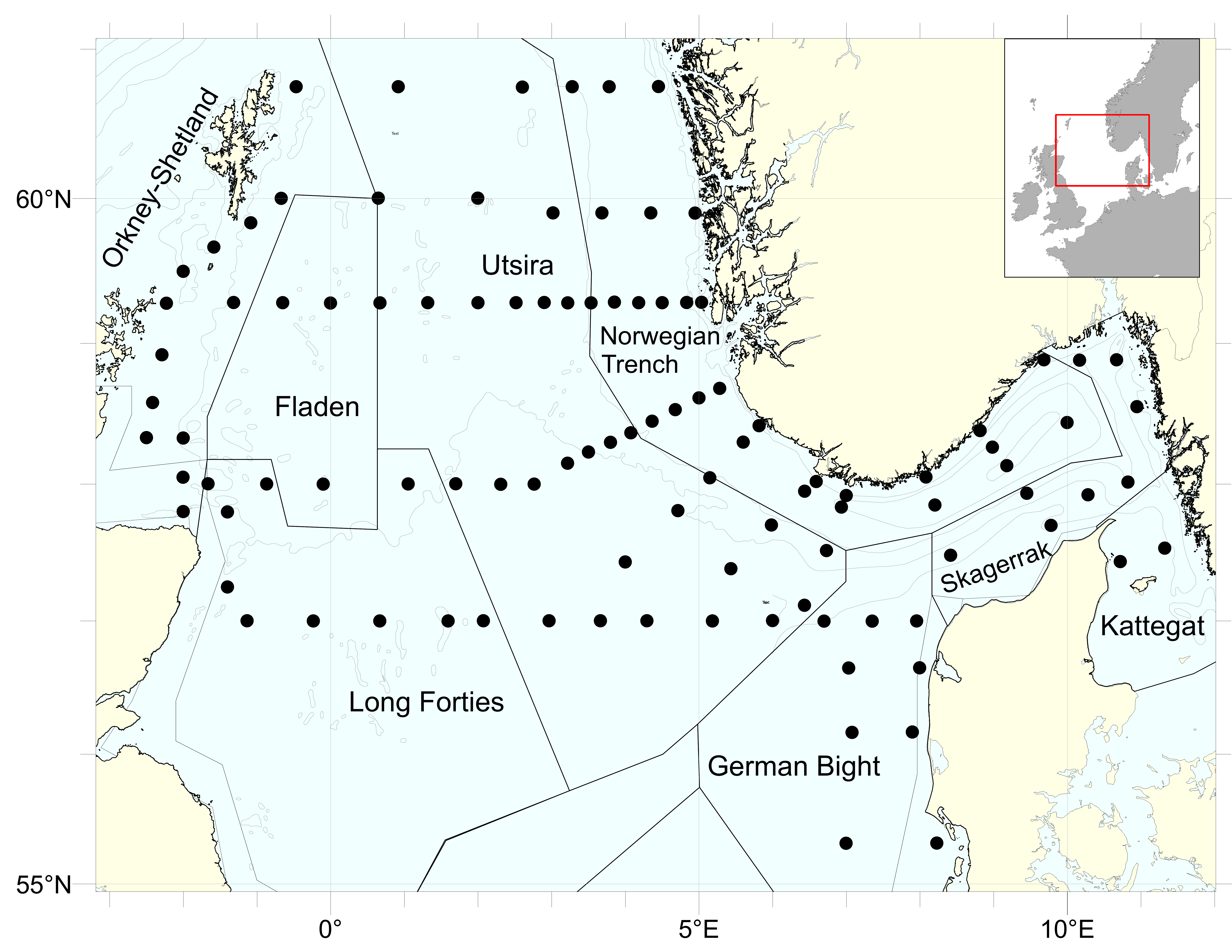}
	\end{center}
	\vspace{-0.4cm}
		\caption{\small Zooplankton sampling areas in the North Sea, including Orkney-Shetland, Fladen, Utsira, Long Forties, German Bight, Skagerrak, Kattegat, and Norwegian Trench, surveyed as part of the climate and plankton monitoring program conducted by the Institute of Marine Research.}
	\label{fig:area_NS}
\end{figure}

In this subsection, we analyze the total biomass data (dry weight) of zooplankton between the years 2006 and 2023 in the North Sea (Figure \ref{fig:area_NS}).
This dataset was collected by the Institute of Marine Research (IMR), Norway, as part of the monitoring program of climate and plankton in the North Sea-Skagerrak.
This dataset includes zooplankton biomass from three main transects (Utsira-Orkney, Hanstholm-Aberdeen, and Torungen-Hirtshals) covered several times per year, as well as the large-scale North Sea Ecosystem cruise, conducted in spring from April to May (\citealt{fr24}). 
The dataset includes all seasons and is geographically restricted to the area between latitudes 55 N and 61 N and between longitudes 2.5 W and 11.5 E. 

Zooplankton sampling and sample treatment were made according to the IMR standard procedure (\citealt{ha20}), which includes vertical tows with a WP2 plankton net (0.25 $m^2$ opening, 180 $\mu m$ mesh size (\citealt{un68}) from near the bottom to the surface. Each sample was split into two equal parts with a Motoda splitter (\citealt{mo59}).
One half of the sample was preserved in 4\% formaldehyde and stored for later analysis (data not presented here). 
The other half was used to determine the dry weight biomass of three size fractions by successive sieving the sample through three meshes (2000 $\mu m$, 1000 $\mu m$, and 180 $\mu m$). 
Samples were transferred to preweighed aluminum trays and dried at 65$^{\circ}$C for $>24$ $h$ until at a constant weight. 
The data analyzed here are the total zooplankton biomass for the entire water column, derived as the sum of the three fractions, and expressed as gram dry weight per square meter surface area.

\begin{figure}[t]
	\begin{center}
\includegraphics[width = 12cm]{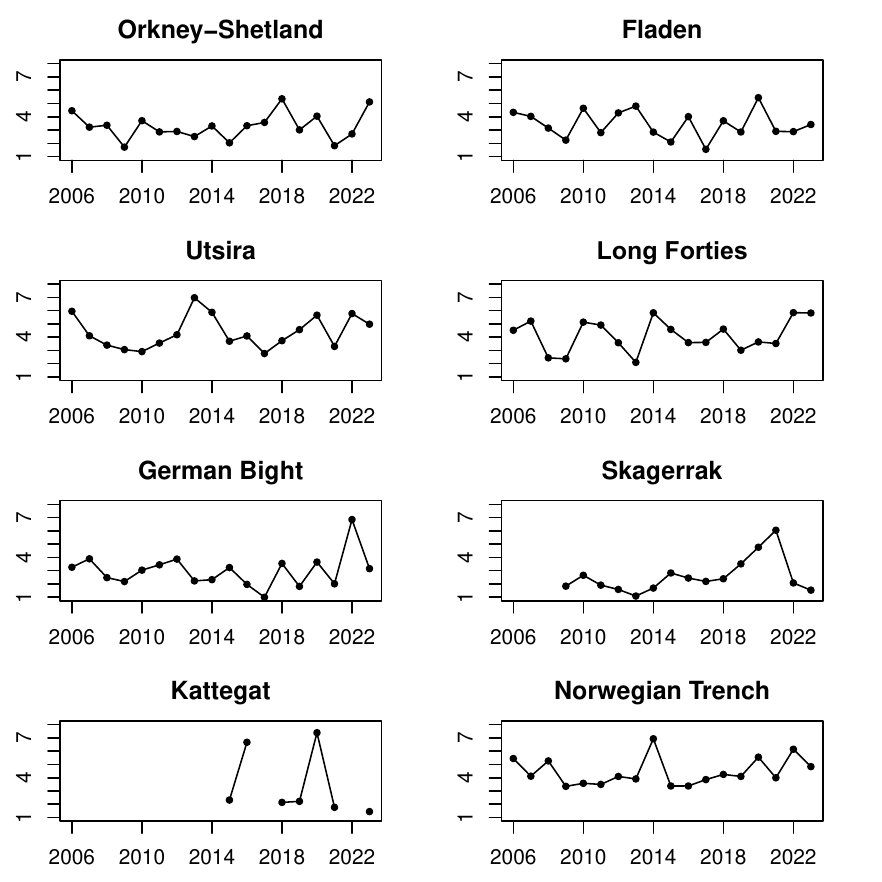}
	\end{center}
	\vspace{-0.4cm}
		\caption{\small Plots of the time series data of zooplankton total biomass for eight sub-areas in the North Sea.}
	\label{fig:plots_Zooplk_IMRsurvey}
\end{figure}

Figure \ref{fig:plots_Zooplk_IMRsurvey} presents the plots of the time series data for the eight areas (Orkney-Shetland, Fladen, Utsira, Long Forties, German Bight, Skagerrak, Kattegat, and Norwegian Trench) shown in Figure \ref{fig:area_NS}.
The data in Kattegat are missing over half of the time points and the data in Skagerrak are missing the first three years.
Therefore, we exclude Kattegat and other areas are used for data from 2009 to 2023.

Our primary objective is the identification of statistically significant regional trends or regional differences concerning the total biomass dataset. 
Therefore, the number of areas and observations are $a=7$ and $n=15$, respectively, and the significance level $\varphi$ is set as 0.05. 
To this end, we carry out the following procedure: for each distinct biomass dataset, we apply our test with $b=6$ to the data from seven areas. 

The abbreviations of area's names are defined as follows:
OSN for Orkeny-Shetland, FG for Fladen, NCNS for Utsira, UKN2 for Long Forties, Ger3 for German Bight, Sk1 for Skagerrak, and NorC for Norwegian Trench. 

We set the hypothesis for applying step 1 in subsection 2.3 as
$$
\text{$H_{0}:$ Mean levels for the overall time series are homogeneous for all areas.}
$$
The overall sample mean supporting homogeneity is 3.56 $g/m^{2}$, and this hypothesis is rejected ($p$-value 0).

Next, step 2 identifies two groups: Group 1 includes Sk1, Ger3, OSN, and FG, and Group 2 includes UKN2, NorC, and NCNS.

In step 3, we set the hypothesis for each group as
$$
\text{$H_{0}:$ Mean levels for the overall time series in Group 1 are homogeneous for all areas.}
$$
$$
\text{$H_{0}:$ Mean levels for the overall time series in Group 2 are homogeneous for all areas.}
$$
The sample mean of the four areas in Group 1 is 3.02 $g/m^{2}$, and the sample mean of the three areas in Group 2 is 4.27 $g/m^{2}$.
For these means, the null hypothesis for each group is not rejected ($p$-value 0.7 for each group).

\begin{figure}[tbp]
	\begin{center}
\includegraphics[width = 12cm]{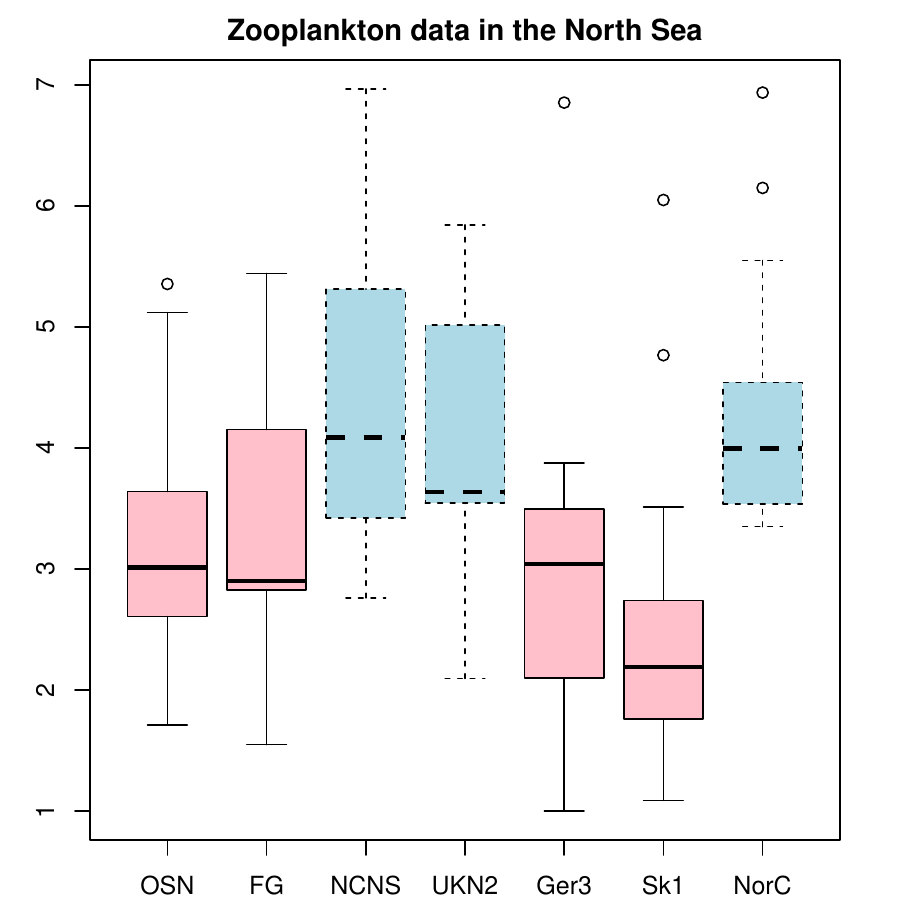}
	\end{center}
	\vspace{-0.4cm}
		\caption{\small Boxplot of the time series data of total biomass from 2009 to 2023 for each sub-area in the North Sea. The abbreviations listed on the horizontal correspond to: OSN for Orkney-Shetland, FG for Fladen, NCNS for Utsira, UKN2 for Long Forties, Ger3 for German Bight, Sk1 for Skagerrak, and NorC for Norwegian Trench. The vertical axis indicates biomass [$g/m^{2}$]. The boxes with solid and dashed borders indicate two different groups based on the output of the method proposed in Section \ref{sec:2.3}.}
	\label{fig:boxplot_IMRSurvey}
\end{figure}

The sample means in the groups are 2.57 $g/m^{2}$ for Sk1, 2.95 $g/m^{2}$ for Ger3, 3.20 $g/m^{2}$ for OSN, 3.37 $g/m^{2}$ for FG, 4.14 $g/m^{2}$ for UKN2, 4.32 $g/m^{2}$ for NorC, and 4.34 $g/m^{2}$ for NCNS.
{Figure} \ref{fig:boxplot_IMRSurvey} shows the boxplot of the time series data for each area.
The boxes with solid and dashed borders indicate two different groups, based on the output from step 2.

The zooplankton biomass presented here represents the vertically integrated abundance for the entire water column, which is affected by the water depth.
Consequently, the shallow areas OSN and FG contain less biomass compared with the deeper NorC, where the plankton net is towed through a larger volume of water. 
In addition, Ger3 and Sk1 are shallow areas, which suggests that both areas include less biomass.
Furthermore, deeper areas contain zooplankton communities with larger-sized zooplankton, as seen in NorC, NCNS, and UKN2.
From ANOVATS, it is verified that this pattern is consistent with previous studies where significant differences in abundance of zooplankton across various regions in the North Sea have been documented (\citealt{fr91}, \citealt{kr03}).
Physical conditions such as ocean currents, temperature, and bottom topography are major drivers for shaping zooplankton distributions and species composition in the North Sea. 
Areas in the central North Sea and NorC tend to have higher zooplankton concentrations owing to the continuous supply of Atlantic water transporting large copepods such as {\it Calanus finmarchicus} into the area from the north (\citealt{he99}, \citealt{ga21}).
In contrast, the shallow areas in Sk1 and Ger3 are dominated by small short-lived plankton forms (\citealt{kr03}).

\subsection{Zooplankton at four sites in Skagerrak}
\begin{figure}[tbp]
	\begin{center}
\includegraphics[width = 12cm]{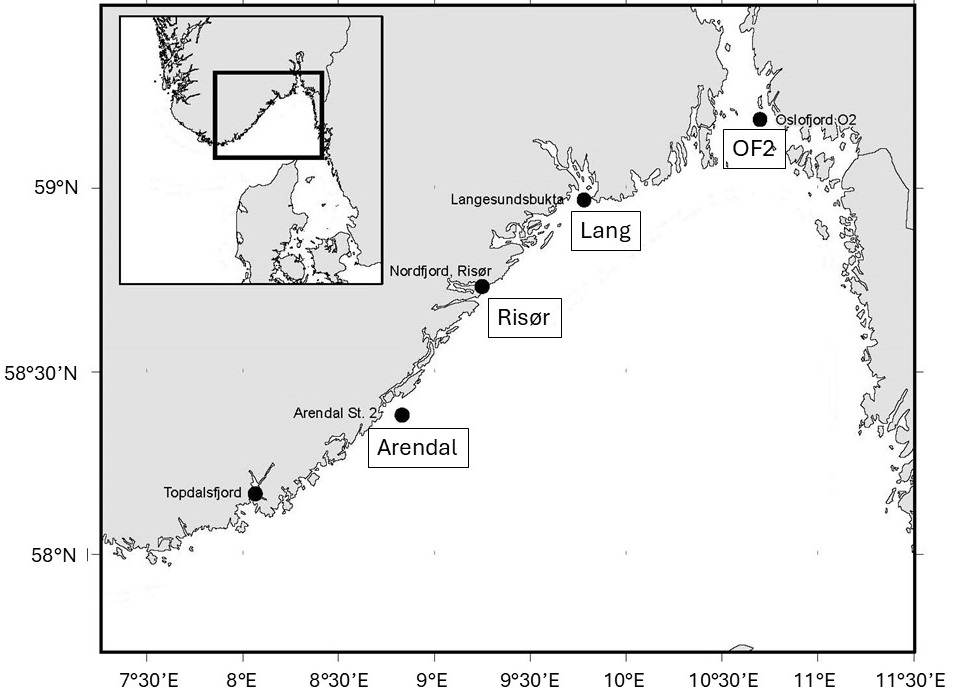}
	\end{center}
	\vspace{-0.4cm}
		\caption{\small 
        Zooplankton sampling areas in Skagerrak, part of the North Sea, including Oslofjord (OF2), Langesund (Lang), Risør (Risør), and Arendal (Arendal), surveyed as part of the Coastal Monitoring program conducted by the Institute of Marine Research. Zooplankton biomass data are believed to exhibit variations in size.}
	\label{fig:area_biomass}
\end{figure}

In this subsection, we apply our method to zooplankton biomass data collected monthly by the IMR from 2013 to 2021 at four coastal sites (Arendal, Langesund, OF2, Ris\o{}r) in Skargerrak (Figure \ref{fig:area_biomass}) as a part of the IMR Coastal Monitoring program.
While the data include a larger number of time points compared with the time series shown in {Section 4.1}, there are missing values in some months in some years. 
For this reason, we have divided the data into four seasons: spring (March, April, May), summer (June, July, August), autumn (September, October, November), and winter (January, February, and December of the previous year) and used quarterly data.
Because there are still missing values in autumn 2019 for Ris\o{}r and Langesund, and in spring 2014 and autumn 2019 for OF2, those data are applied by Box-Cox transformation and imputing the parts by an autoregressive model. 
This procedure is conducted by the functions {\it boxcox} and {\it arfit} in R package {\it TSSS} (\citealt{ki23}) and R code to impute the missing values,  written based on the theory of state space modeling (\citealt{ki21}).
The final time series includes 36 quarterly points.

\begin{figure}[ht]
	\begin{center}
\includegraphics[width = 12cm]{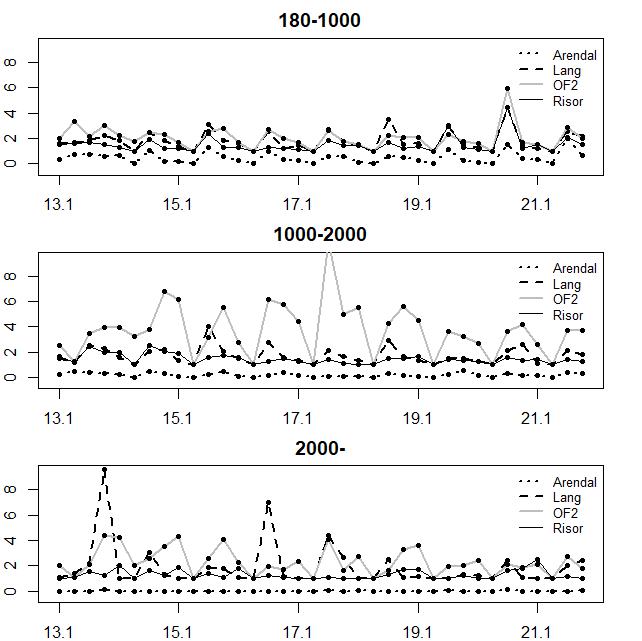}
	\end{center}
	\vspace{-0.4cm}
		\caption{\small Plots for the quarterly time series data of zooplankton biomass for each size group (180 $\mu m$ - 1000 $\mu m$, $1000 \mu m$ - $2000 \mu m$, and $>2000$ $\mu m$), sampled at four sites in Skagerrak.}
	\label{fig:plots_bio}
\end{figure}

Zooplankton sampling and sample treatment followed the IMR standard procedure (\citealt{ha20}) as described in Section 4.1. 
The dataset is given as grams dry weight per square meter surface area ($g/m^2$) and grouped into three categories according to size fractions: 180 $\mu m$ - 1000 $\mu m$, $1000 \mu m$ - $2000 \mu m$, and $>2000$ $\mu m$.
The time series data for each size fraction are presented in Figure \ref{fig:plots_bio}. 

\begin{figure}[bp]
	\begin{center}
\includegraphics[width = 12cm, page=1]{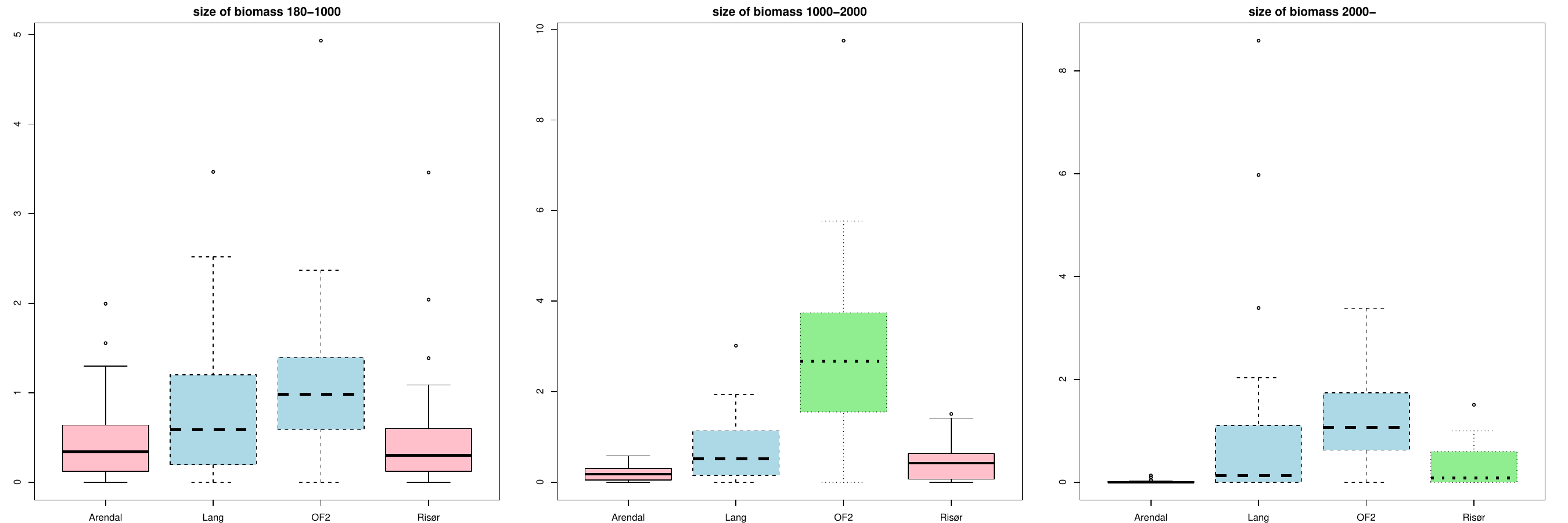}
	\end{center}
	\vspace{-0.4cm}
		\caption{\small Boxplots for the biomass data.  {The horizontal corresponds to sites.} The boxplots with solid borders are the subareas for which the null hypothesis was rejected by ANOVATS. The  plots with dashed and dotted borders are areas that were divided into two subgroups within a group for which the null hypothesis was rejected. The boxplot for Arendal with biomass size $>2000$ $\mu m$ has a box with a solid border.}
	\label{fig:boxplots_bio}
\end{figure}

We apply our procedure explained in {Section 2.2} with $b=8$ and the result is given in Figure \ref{fig:boxplots_bio}.

In step 1, the overall sample mean levels supporting the null hypothesis are 0.705 $g/m^2$ for size fractions 180 $\mu m$ - 1000 $\mu m$, 1.01 $g/m^2$ for  size fractions 1000 $\mu m$ - 2000 $\mu m$, and 0.619 $g/m^2$ for size fractions  $>2000$ $\mu m$.
The null hypothesis is rejected for all size fractions (p-value 0 for each).
Furthermore, step 2 divides the sites into two groups as follows:
\begin{enumerate}
  \item[] \textbf{180–1000 $\mu$m}: 
    Group 1: Arendal, Risør; 
    Group 2: Lang, OF2
  \item[] \textbf{1000–2000 $\mu$m}: 
    Group 1: Arendal, Risør, Lang; 
    Group 2: OF2
  \item[] \textbf{$>$2000 $\mu$m}: 
    Group 1: Arendal, Risør; 
    Group 2: Lang, OF2.
\end{enumerate}
For 180 $\mu m$ - 1000 $\mu m$, step 3 does not further subdivide Group 1 (p-value 0.86, sample mean for null 0.488 $g/m^2$) or Group 2 (p-value 0.069, sample mean for null 0.921 $g/m^2$).

For the case of 1000 $\mu m$ - 2000 $\mu m$, step 3 subdivides Group 1 into two groups: Group 1-1 includes Arendal and Ris\o{}r, and Group 1-2 includes  Lang (p-value 0, sample mean for null 0.447 $g/m^{2}$).
Step 3 does not further subdivide Group 1-1 (p-value 0.14, sample mean for null 0.322 $g/m^2$).

Finally, for $>2000$ $\mu m$, step 3 subdivides Group 1 into two groups, one with Arendal and the other with Ris\o{}r (p-value 0, sample mean for null 0.151 $g/m^{2}$).
However, Group 2 is not subdivided into two groups by step 3 (p-value 0.24, sample mean for null 1.09 $g/m^{2}$).

Table \ref{tab:sample_mean} summarizes the sample mean for each site according to size fractions.
\begin{table}[t]
 \caption{Sample means for each site according to different size fractions}
 \label{tab:sample_mean}
 \centering
  \begin{tabular}{ccc}\hline
size fractions [$\mu m$] & sites & sample mean [$g/m^{2}$]\\\hline
$180 - 1000$ & Arendal & 0.480\\
 & Langesund & 0.800\\
 & OF2 & 1.04\\
 & Ris\o{}r & 0.497\\
 \hline
 $1000 - 2000$ & Arendal & 0.198\\
 & Langesund & 0.698\\
 & OF2 & 2.69\\
 & Ris\o{}r & 0.446\\
 \hline
 $>2000$ & Arendal & 0.011\\
 & Langesund & 0.886\\
 & OF2 & 1.289\\
 & Ris\o{}r & 0.289\\
 \end{tabular}
\end{table}
The biomass data sampled at Langesund and OF2 are identified in the same group for each size fraction. 
The data from Ris\o{}r is divided into subgroups only for size fraction $>2000$ $g/m^2$, and the data from Arendal remains undivided for all size fractions.
These results can be explained from the point view of the sampling depth and the position of the sampling station in relation to the coast.
The sampling depths and the positions of the sampling stations are provided in Table \ref{tab:depth_location}.
\begin{table}[ht]
 \caption{Sampling depths and positions of sampling stations}
 \label{tab:depth_location}
 \centering
  \begin{tabular}{ccc}\hline
sites & sampling depth & position\\\hline
Arendal & 50 $m$ & in the coastal current\\
Langesund & 200 $m$ & outside the fjord sill\\
OF2 & 350 $m$ & outside the fjord in the coastal current\\
Ris\o{}r & 160 $m$ & inside the fjord sill\\
 \end{tabular}
\end{table}
Sampling sites with deep sampling depths (OF2 and Langesund) usually have larger total biomass values.
This is explained by the plankton net being towed through a larger sampling volume from the bottom to the surface. 
Furthermore, deeper water layers often harbor larger zooplankton species (\citealt{ak89}), contributing to higher biomass values. 
This is confirmed by the high proportion of the size fraction $>2000$ $\mu m$ at these sites.
Ris\o{}r is located inside the fjord sill and may have a somewhat different zooplankton community. 
The presence of a sill at the mouth of the fjord serves as a barrier, limiting  water exchange between the fjord and the adjacent coastal current, and generally the zooplankton populations of the fjord differ significantly from those of open ocean waters, with a decrease in the abundance of zooplankton towards the inner regions of the fjord (\citealt{sa95}).
The boxplots of the data are shown in Figure \ref{fig:boxplots_bio}. 
The boxplots with solid borders correspond to the sites for which the null hypothesis was rejected.
The boxplots clearly show the grouped areas have larger means than those of the rejected areas from ANOVATS.
The results are supported by the following facts: Langesund and OF2 differ in depth and species composition.
OF2 has a higher proportion of large zooplankton, and it is probably affected more by advection of water masses from NorC. 
OF2 is deeper (with a basin) and thus a better habitat for large plankton like euphausiids and copepods.

\vspace{1cm}

These results estimated by ANOVATS are supported by multiple sources in the literature and are consistent with biological/ecological facts. 
In the field of marine ecology, data analysis using ANOVATS represents the first attempt to statistically prove these facts.
Variations in marine resource abundance based on regional differences are presented within a simple visualization and descriptive statistical framework not only for the North Sea but also for the Barents Sea (\citealt{sr25})  and the Norwegian Sea (\citealt{ic25}). Data analysis using ANOVATS must be pursued to establish the statistical credibility of these variations.

\section{Conclusion}
{We provided a subsampling-based test, ANOVATS (ANOVA for small-sample time series data), for exploring regional differences. 
While the existing ANOVA method for time series data requires estimation of spectral density, which is not appropriate for small-sample time series data, our proposed method avoids spectral estimation and achieves better size control, even for small-sample time series data.
Time series data often contains trend and/or seasonality.
Our approach is extend to take these properties into account.

There are only a few decades of annual data available for studying the impact of climate change on the diverse species of marine resource and environmental changes caused by human activity in IEA.
The proposed method is practical for comparing the mean levels of these changes based on short time series data.
This study shows an example of the application of zooplankton data. 
The differences among strata are attributable to topographical factors, and although they are shown qualitatively, it has become possible to show statistically significant differences without prior information using the proposed method.
{The fact that the output from our proposed method is supported by previous studies also indicates that this method can be used to verify whether the data contain any artifacts.}
The North Sea is considered to be particularly vulnerable to oil activities in the Northern Sea area (\citealt{tw24}), but there are many fish groups in each strata that should be investigated for their biological production, diversity,  and ecosystems. This information is stored in multiple databases, such as the trawl surveys database by ICES (\citealt{ic23}) and an indicator assessment based on the Oslo and Paris Commissions (\citealt{OS22}). 
Investigations on ecosystem status based on subregions are also conducted in several sea areas (e.g., \citealt{ic22}), and it is possible to apply our approach to the differences between ecological regions to investigate the biodiversity for each species. 
The approach we proposed has potential to be a standard method within the IEA. 
By adding quantitative comparisons of time series data obtained from ANOVATS, the outcomes are incorporated into risk assessments in IEA, enabling more reliable implementation of ecosystem-based management.

\begin{acks}[Acknowledgments]
We would like to thank Prof.\ Li Wang and Prof.\ Xingzhong Xu for kindly sharing a copy of their paper with us.
We also thank the captains and crews of the various research vessels of the Institute of Marine Research, as well as the many people involved in collecting and processing the data here analyzed.
We are grateful to Edanz (\url{https://jp.edanz.com/ac}) for editing the English text of a draft of this manuscript.
We used ChatGPT (based on GPT-3.5, developed by OpenAI) for the purposes of language improvement and coding assistance.
\end{acks}

\begin{funding}
The first author gratefully acknowledge JSPS Grant-in-Aid for Early-Career Scientists JP23K16851 and Research Fellowship Promoting International Collaboration of the Mathematical Society of Japan.
\end{funding}

\bibliographystyle{imsart-nameyear}
\bibliography{ref}

\appendix

\section{Simulation study}\label{sec:A}

\subsection{Data generating process}\label{sec:dgp}
To evaluate finite sample performance of the proposed test, we consider two cases: 
Cases 1 and 2 correspond to independent and correlated groups, respectively. For each case, we use Processes 1--4, which were considered in \cite{galt2022}, as disturbance processes of the one-way random effects model defined in (1) with $p=1$.
For both Cases 1 and 2, Processes 1--3 are defined as the moving average model of order 1 processes $\bm{e}_{t} = \bm{\nu}_{t} + {\bm \Psi} \bm{\nu}_{t-1},$ where ${\boldsymbol e}_{t}=({\bm e}_{1t}^\top,\ldots,{\bm e}_{at}^\top)^\top$, and Process 4 is defined as the GARCH process \[
e_{it} = h_{it}^{1/2} \nu_{it}, \quad i = 1, \dots, a, \quad 
\begin{pmatrix}
h_{1t} \\
\vdots \\
h_{at}
\end{pmatrix} 
= 
\begin{pmatrix}
1 \\
\vdots \\
1 \\
\end{pmatrix}
+
0.1{\bm \Psi}
\begin{pmatrix}
e_{1t}^2 \\
\vdots \\
e_{at}^2 \\
\end{pmatrix}
+
\begin{pmatrix}
0.1 h_{1, t-1} \\
\vdots \\
0.1 h_{a, t-1}
\end{pmatrix}.
\]
{
\begin{enumerate}
    \item [Case 1]
Let ${\bm \Psi}$ be $0.5\bm I_a$.
Each component of $\bm{\nu}_t$, which is independent of the other components, follows the standard normal for Processes 1 and 4, the centered t-distribution with five degrees of freedom for Process 2, and the centered skew normal distribution with location 0, scale 1, and shape 50  for Process 3.
    \item [Case 2]
Let ${\bm \Psi}$ be a block diagonal matrix whose block diagonal matrices are given by
\begin{align*}
\begin{pmatrix}
0.7&0&0\\
0&-0.5&0\\
0.3&0.1&0.3\\
\end{pmatrix}
\end{align*}
and  let
$\bm\Sigma^{\bm{\nu}}=(\Sigma^{\bm{\nu}}_{ij})_{i,j=1,\ldots,a}$ be an $a$-by-$a$ covariance matrix such that $\Sigma^{\bm{\nu}}_{ii}=1$ and $\Sigma^{\bm{\nu}}_{j(j+1)}=\Sigma^{\bm{\nu}}_{(j+1)j}=0.5$ for $i\in\{1,\ldots,a\}$ and $j\in\{1,\ldots,a-1\}$.
The process $\{\bm{\nu}_t\}$ follows the centered multivariate distribution with covariance matrix $\Sigma^{\bm{\nu}}$ for Processes 1 and 4, the centered multivariate t-distribution with five degrees of freedom and scale matrix $\Sigma^{\bm{\nu}}$ for Process 2, the centered multivariate skew normal distribution with location ${\bm 0_a}$, correlation matrix $\Sigma^{\bm{\nu}}$, and shape $50{\bm 1}_a$, where  ${\bm 0_a}$ and ${\bm 1}_a$ are a-dimensional vectors whose components are all 0 and 1, respectively, for Process 3.
\end{enumerate}
The MA model is indeed one of the simplest time series models, capable of capturing short-term dependencies and providing easily understandable simulation results. Conversely, the GARCH model is more sophisticated and specifically designed to address conditional heteroskedasticity.}


\subsection{Empirical size}
{To investigate the empirical size of our test, we set the number of groups as $a=3,9,15$, sample size as $n=20,30,50,70,100$, and subsampling block length $b=\lfloor c n^{1/3} \rfloor$, where $c\in\{1,1.5,2,2.5,3,4,5,6\}$.
Let $p=1$.
Then for given $a$, $n$, and $b$, generate time series $z_{it}$ defined in (1), where $\mu=0$, ${\psi}_{i}=0$ for all $i$, and ${\bm e}_{t}$ as described in Section and compute the $p$-value. Iterate 200 times and calculate the empirical size with significance level $\varphi=0.05$.}

{Because the constant $c=2.5$ provides the best performance among the considered range, we fix the block length as $b=\lfloor 2.5 n^{1/3} \rfloor$.
Figure \ref{fig:null} shows the empirical sizes. The horizontal of the subplots corresponds to the time series length.
The empirical size tends to get closer to $0.05$ as the time series length increases. Even for small time series lengths, the performance of our test is reasonably good. 
From Figures 1 and 2 of \cite{galt2022suppl}, we know the classical tests based on $S_n^\prime$ and $S_n^{\prime\prime}$ exhibit size distortion owing to between-area correlations and the number of areas, respectively, even for $n=1000$ and $n=2000$. As expected, our proposed test shows good size control for small-sample time series.}
Apparently, there is no significant difference in the results for Cases 1 and 2.

\subsection{Empirical power}
Next, we investigate the empirical power of our test. Setting $a=6$, $p=1$,
sample sizes $n\in\{20,30,50,70,100\}$, subsampling block length $b=\lfloor 2.5 n^{1/3} \rfloor$, we generate time series $z_{it}$ defined in (1), where $\mu=0$, $(\psi_1,\ldots,\psi_6):=(0,0,0,1,1,1)^\top$, and ${\bm e}_{t}$ as described in Section \ref{sec:dgp}. There are two clusters in this setting. Then we apply our method described in Section 2.3 with significance level $\varphi=0.05$ and repeat 200 times. Let ${\rm Area}_i$ denote the area name of $\{z_{it}\}$.

The first and second rows of Figure \ref{fig:alt} correspond to Cases 1 and 2, respectively.

The first column of Figure \ref{fig:alt} shows  the empirical probability of rejecting $H_0:\psi_1=\cdots=\psi_6$ and dividing the areas $({\rm Area}_1,\ldots,{\rm Area}_6)$ into two groups $({\rm Area}_1,\ldots,{\rm Area}_3)$ and $({\rm Area}_4,\ldots,{\rm Area}_6)$.
Our test has power even with a small time series length $n$, although the power is small when $n=20$. This is attributable to the variance in the time series and the degree of separation between clusters. We see that the t-distribution provides a smaller power owing to the large fluctuation, which hides the regional effect.

The second column shows, for the data that were correctly divided in the first step, the empirical size for not rejecting the null hypothesis for both groups.
The third column shows  the number of cases for which the areas were divided correctly in the first step.
As $n$ increases, the size tends to approach or slightly exceed 0.05. When 
$n$ is small, the size is less than 0.05, resulting in fewer trials and smaller observed sizes.

The fourth column shows the  empirical probability of reaching the correct clustering results. When $n$ is relatively large, correct clustering results are achieved. We observe that the empirical power for Case 2 is larger than that for Case 1.

\begin{figure}[htbp]
	\begin{center}
\includegraphics[width = 12cm]{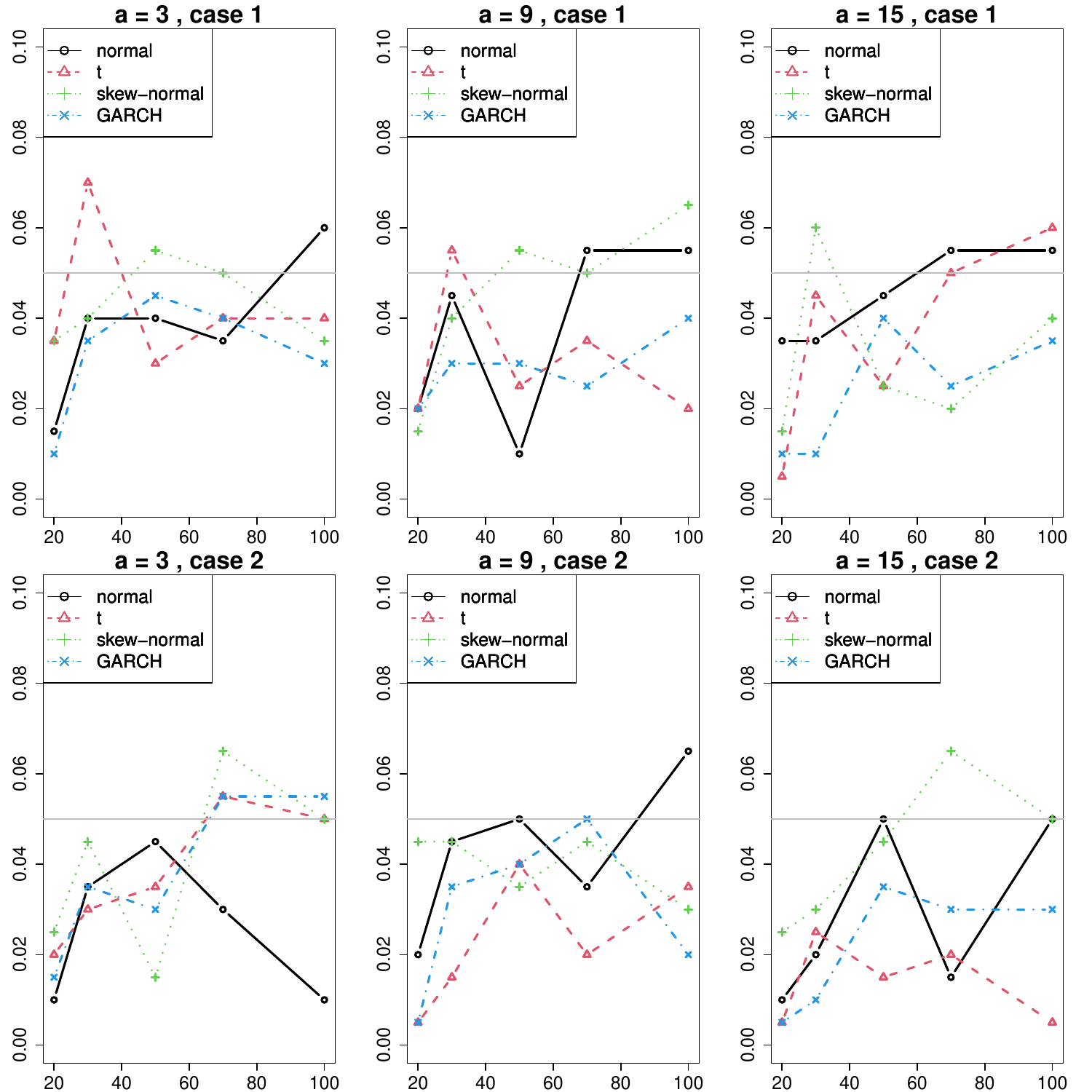}
	\end{center}
	\vspace{-0.4cm}
		\caption{\small Empirical size of the proposed test. The horizontal and vertical axes correspond to time series length $n$ and empirical size, respectively. 
The left, middle, and right columns correspond to $a=3,9,15$, respectively.
The top and bottom rows correspond to Cases 1 and 2, respectively.}
	\label{fig:null}
\end{figure}

\begin{figure}[htbp]
	\begin{center}
\includegraphics[width = 10cm]{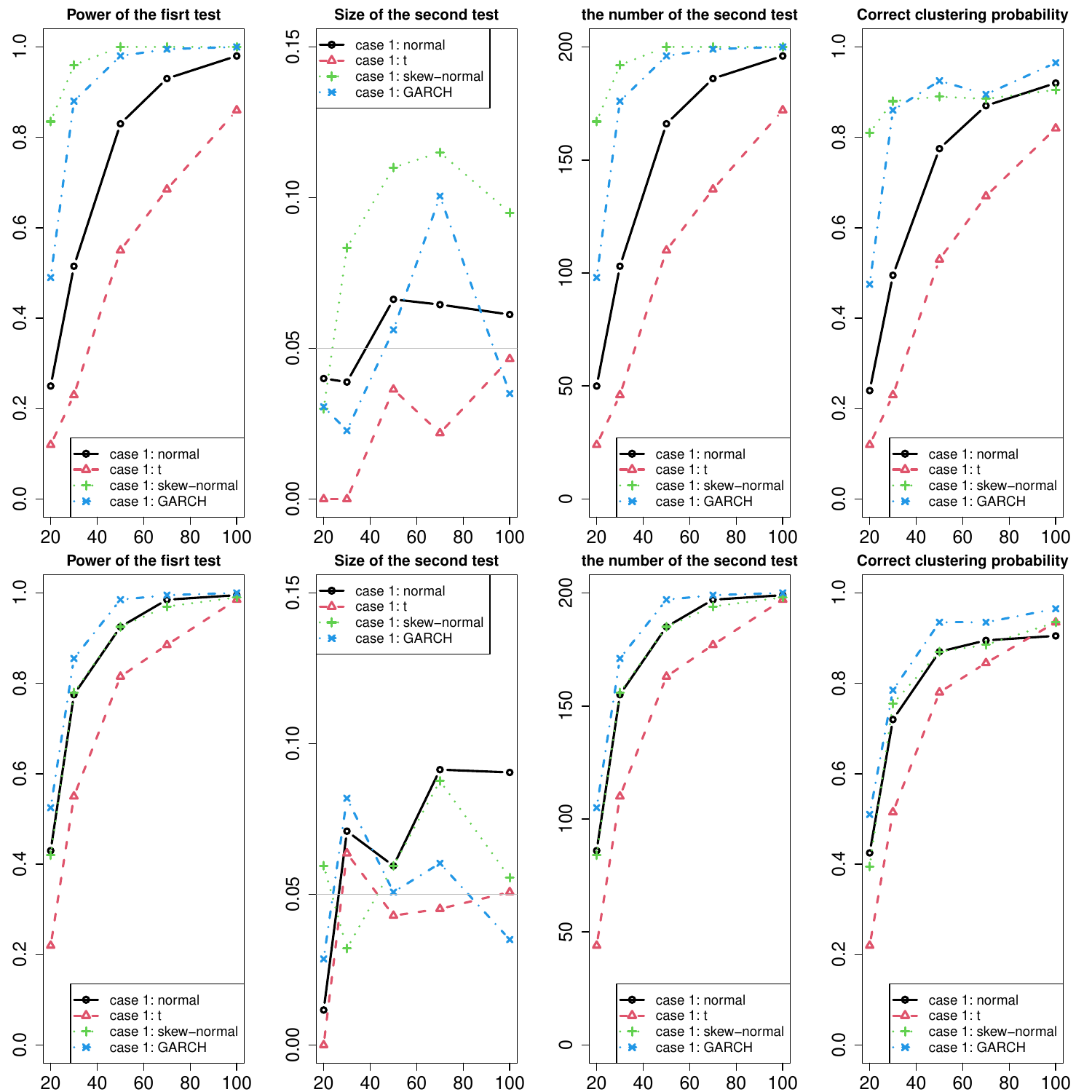}
	\end{center}
	\vspace{-0.4cm}
		\caption{\small The empirical probability of rejecting the null and dividing the areas correctly in the first step (first column),
the empirical size for not rejecting the null hypothesis for both groups for data that was correctly divided in the first step (second column),
the number of cases for which the areas were divided correctly in the first step (third column), and
the empirical probability of reaching the correct clustering results (fourth column).
The horizontal and vertical axes correspond to time series length $n$ and probability, respectively. The top and bottom rows correspond to Cases 1 and 2, respectively.}
	\label{fig:alt}
\end{figure}

\clearpage

\section{Proofs}\label{sec:B}
\subsection{Proof of Theorem 2.1}

This theorem can be proved in the same way as \citet[Proof of Theorem 4.2]{gkvvdh22}. The essential tool is \citet[Proposition 7.3.1]{pr99}, which implies that 
$\sup_{x\in\mathbb R}\left|H_{n,b}(x)-H(x)\right|=o_p(1)$,
where 
$$H_{n,b}(x):=\frac{1}{n-b+1}\sum_{t=1}^{n-b+1}\mathbb I\left\{\frac{b}{1-\frac{b}{n}}\sum_{i=1}^{a}\skakko{{\boldsymbol e}_{i.,b,t}-{\boldsymbol e}_{..,b,t}}^\top\skakko{{\boldsymbol e}_{i.,b,t}-{\boldsymbol e}_{..,b,t}}\leq x\right\}$$ and $H$ is the c.d.f. of the asymptotic distribution of $T_n$ under the null with
${{\boldsymbol e}_{i.,b,t}}=\sum_{j=t}^{t+b-1}{\boldsymbol e}_{ij}/{{b}}$ and 
${{\boldsymbol e}_{..,b,t}}=\sum_{i=1}^{a}\sum_{j=t}^{t+b-1}{\boldsymbol e}_{ij}/{(a{b})}$. This yields that  the proposed test has the asymptotic size $\varphi$.

Next, we show the consistency of the test. Note that
\begin{align*}
T_{n}=&
n\sum_{i=1}^{a}
\skakko{\boldsymbol {{\psi}}_{i}-\boldsymbol {{\psi}}_{.}
+{\boldsymbol e}_{i.}-{\boldsymbol e}_{..}}^\top
\skakko{\boldsymbol {{\psi}}_{i}-\boldsymbol {{\psi}}_{.}
+{\boldsymbol e}_{i.}-{\boldsymbol e}_{..}}\\
=&
n\sum_{i=1}^{a}\skakko{\boldsymbol {{\psi}}_{i}-\boldsymbol {{\psi}}_{.}}^\top\skakko{\boldsymbol {{\psi}}_{i}-\boldsymbol {{\psi}}_{.}}
+2n\sum_{i=1}^{a}\skakko{\boldsymbol {{\psi}}_{i}-\boldsymbol {{\psi}}_{.}}^\top\skakko{{\boldsymbol e}_{i.}-{\boldsymbol e}_{..}}\\
&
+n\sum_{i=1}^{a}\skakko{{\boldsymbol e}_{i.}-{\boldsymbol e}_{..}}^\top\skakko{{\boldsymbol e}_{i.}-{\boldsymbol e}_{..}}\\
=&n\skakko{\zeta_{\psi,\psi}+2\zeta_{\psi,e_n}+\zeta_{e_n,e_n}}
\end{align*}
and
\begin{align*}
T_{n,b,t}=&\frac{b}{1-\frac{b}{n}}\skakko{\zeta_{\psi,\psi}+2\zeta_{\psi,e_{b,t}}+\zeta_{e_{b,t},e_{b,t}}},
\end{align*}
where
\begin{align*}
\zeta_{\psi,\psi}:=&\sum_{i=1}^{a}\skakko{\boldsymbol {{\psi}}_{i}-\boldsymbol {{\psi}}_{.}}^\top\skakko{\boldsymbol {{\psi}}_{i}-\boldsymbol {{\psi}}_{.}},\quad
\zeta_{\psi,e_n}:=\sum_{i=1}^{a}\skakko{\boldsymbol {{\psi}}_{i}-\boldsymbol {{\psi}}_{.}}^\top\skakko{{\boldsymbol e}_{i.}-{\boldsymbol e}_{..}}\\
\zeta_{e_n,e_n}:=&\sum_{i=1}^{a}\skakko{{\boldsymbol e}_{i.}-{\boldsymbol e}_{..}}^\top\skakko{{\boldsymbol e}_{i.}-{\boldsymbol e}_{..}},\quad
\zeta_{\psi,e_{b,t}}:=\sum_{i=1}^{a}\skakko{\boldsymbol {{\psi}}_{i}-\boldsymbol {{\psi}}_{.}}^\top\skakko{{\boldsymbol e}_{i.,b,t}-{\boldsymbol e}_{..,b,t}},\\\text{ and }
\zeta_{e_{b,t},e_{b,t}}:=&\sum_{i=1}^{a}\skakko{{\boldsymbol e}_{i.,b,t}-{\boldsymbol e}_{..,b,t}}^\top\skakko{{\boldsymbol e}_{i.,b,t}-{\boldsymbol e}_{..,b,t}}
\end{align*}
with
${{\boldsymbol e}_{i.,b,t}}=\sum_{j=t}^{t+b-1}{\boldsymbol e}_{ij}/{{b}}$, 
${{\boldsymbol e}_{..,b,t}}=\sum_{i=1}^{a}\sum_{j=t}^{t+b-1}{\boldsymbol e}_{ij}/{(a{b})}$. It holds that
\begin{align*}
p_n
&=
\frac{1}{n-b+1}\sum_{t=1}^{n-b+1}\mathbb I\{\left|T_{n,b,t}\right|> \left|T_{n}\right|\}\\
&\leq
\frac{1}{n-b+1}\\
&\times\sum_{t=1}^{n-b+1}\mathbb I\left\{\frac{b}{1-\frac{b}{n}}\left|\zeta_{\psi,\psi}+2\zeta_{\psi,e_{b,t}}\right|+\frac{b}{1-\frac{b}{n}}\left|\zeta_{e_{b,t},e_{b,t}}\right|> n\left|\zeta_{\psi,\psi}+2\zeta_{\psi,e_n}\right|-n\left|\zeta_{e_n,e_n}\right|\right\}\\
&=1- H_{n,b}\skakko{n\left|\zeta_{\psi,\psi}+2\zeta_{\psi,e_n}\right|-\frac{b}{1-\frac{b}{n}}\left|\zeta_{\psi,\psi}+2\zeta_{\psi,e_{b,t}}\right|-n\left|\zeta_{e_n,e_n}\right|}\\
&=1- H\skakko{n\left|\zeta_{\psi,\psi}+2\zeta_{\psi,e_n}\right|-\frac{b}{1-\frac{b}{n}}\left|\zeta_{\psi,\psi}+2\zeta_{\psi,e_{b,t}}\right|-n\left|\zeta_{e_n,e_n}\right|}+o_p(1).
\end{align*}
Since $\zeta_{\psi,e_n}=O_p(1/\sqrt n)$, $\zeta_{\psi,e_{b,t}}=O_p(1/\sqrt b)$, and $\zeta_{e_n,e_n}=O_p(1/n)$, 
there exists $M>0$ such that for all $n\in\mathbb N$,
\begin{align*}
{\rm P}\skakko{\sqrt n\left|2\zeta_{\psi,e_n}\right|>M}<\epsilon,\quad
{\rm P}\skakko{\sqrt b\left|2\zeta_{\psi,e_{b,t}}\right|>M}<\epsilon,\text{ and }
{\rm P}\skakko{n\left|\zeta_{e_n,e_n}\right|>M}<\epsilon.
\end{align*}
Therefore, we obtain, for any $\epsilon^\prime>0$,
\begin{align*}\label{proof:consistency}
&{\rm P}\skakko{p_n>\epsilon^\prime}\\
\leq
&{\rm P}\skakko{1- H\skakko{n\left|\zeta_{\psi,\psi}+2\zeta_{\psi,e_n}\right|-\frac{b}{1-\frac{b}{n}}\left|\zeta_{\psi,\psi}+2\zeta_{\psi,e_{b,t}}\right|-n\left|\zeta_{e_n,e_n}\right|}>\epsilon^\prime}+o(1)\\
\leq
&{\rm P}\skakko{1- H\skakko{n\zeta_{\psi,\psi}-n\left|2\zeta_{\psi,e_n}\right|-\frac{b}{1-\frac{b}{n}}\left|\zeta_{\psi,\psi}\right|-\frac{b}{1-\frac{b}{n}}\left|2\zeta_{\psi,e_{b,t}}\right|-n\left|\zeta_{e_n,e_n}\right|}>\epsilon^\prime}+o(1)\\
\leq
&{\rm P}\skakko{1- H\skakko{\skakko{n-\frac{b}{1-\frac{b}{n}}}\zeta_{\psi,\psi}-\skakko{\sqrt n+\frac{\sqrt b}{1-\frac{b}{n}}+1}M}>\epsilon^\prime}+3\epsilon+o(1).
\end{align*}
Since there exists $N\in\mathbb N$ such that for $n\geq N$, 
$${\rm P}\skakko{1- H\skakko{\skakko{n-\frac{b}{1-\frac{b}{n}}}\zeta_{\psi,\psi}-\skakko{\sqrt n+\frac{\sqrt b}{1-\frac{b}{n}}+1}M}>\epsilon^\prime}<\epsilon,$$
we obtain $p_n=o_p(1)$.
\qed

\subsection{Proofs of the convergences in (5)}\label{sec:proof_phi}
From the proof of Theorem 2.1, $p_n$ converges in distribution to a uniform distribution on (0,1) under $H_0$. Also, $p_n - \varphi_n$ converges in distribution to the same, provided $\varphi_n\to0$, and thus ${\rm P}\skakko{p_n<\varphi_n}\to0$ as $n\to\infty$.

Next, we consider the case under $H_1$.
First, we note that Lemma 1 of \cite{il06} yields, for $a\geq2$ and $u>a$, that
\begin{align*}
{\rm P}\skakko{\chi_a^2\geq u}
\leq&
\frac{1}{\sqrt {2\pi}}
\frac{\sqrt a}{u-a+2}
\skakko{\frac{\exp(1)}{a}u}^{\frac{a}{2}}
\exp\skakko{-\frac{u}{2}}.
\end{align*}
Let $c_n$ denote the quantity
\begin{align*}
c_n:=\skakko{n-\frac{b}{1-\frac{b}{n}}}\zeta_{\psi,\psi}-\skakko{\sqrt n+\frac{\sqrt b}{1-\frac{b}{n}}+1}M,
\end{align*}
which is $O(n)$ as $n\to\infty$.
Then, from the proof of Theorem 2.1, we have, under $H_1$, 
\begin{align*}
{\rm P}\skakko{p_n>\varphi_n}
\leq
&{\rm P}\skakko{1- H\skakko{c_n}>\varphi_n}+3\epsilon+o(1)\\
\leq
&{\rm P}\skakko{
\frac{1}{\sqrt {2\pi}}
\frac{\sqrt a}{c_n-a+2}
\skakko{\frac{\exp(1)}{a}c_n}^{\frac{a}{2}}
\exp\skakko{-\frac{c_n}{2}}>\varphi_n}+3\epsilon+o(1),
\end{align*}
which, provided the condition
$
\varphi_n/\skakko{
c_n^{\frac{a}{2}-1}
\exp\skakko{-\frac{c_n}{2}}
} \to\infty
$ as $n\to\infty$, is $o(1)$.\qed


\end{document}